\documentclass[sigconf]{acmart}

\hypersetup{draft}

\setcopyright{acmcopyright}
\copyrightyear{2018}
\acmYear{2021}
\acmDOI{10.1145/1122445.1122456}

\acmConference[Somewhere]{We do not know yet}{..., 2021}{Virtual}
\acmBooktitle{Booktitle}
\acmPrice{15.00}
\acmISBN{978-1-4503-XXXX-X/18/06}

\usepackage{lineno,hyperref}
\usepackage{csquotes} 

\usepackage[]{todonotes}

\usepackage{rotating}

\usepackage{multirow}

\usepackage{subcaption}

\usepackage{paralist}

\usepackage{fontawesome}

\usepackage{tikz}
\usepackage{pgfplots}
\usetikzlibrary{patterns}

\usepackage[capitalise,nameinlink]{cleveref}
\crefname{section}{Sect.}{Sect.}
\Crefname{section}{Section}{Sections}
\crefname{figure}{Fig.}{Figs.}
\Crefname{figure}{Figure}{Figures}
\crefname{table}{Tab.}{Tabs.}
\Crefname{table}{Table}{Tables}
\crefname{lstlisting}{List.}{List.}
\Crefname{lstlisting}{Listing}{Listings}

\newcommand{\eg}{e.\,g.,\ }
\newcommand{\ie}{i.\,e.,\ }

\usepackage[inline]{enumitem}

\usepackage{url}

\begin{document}

\title{GraphMatch: Subgraph Query Processing on FPGAs}

\author{Jonas Dann}
\email{jonas.dann@sap.com}
\affiliation{%
    \institution{Heidelberg University \& SAP}
    \city{Walldorf (Baden)}
    \country{Germany}
}

\author{Tobias Götz}
\email{goetzt@in.tum.de}
\affiliation{%
   \institution{Technical University of Munich \\\& SAP}
    \city{Munich}
    \country{Germany}
}

\author{Daniel Ritter}
\email{daniel.ritter@sap.com}
\affiliation{%
    \institution{SAP}
    \city{Walldorf (Baden)}
    \country{Germany}
}

\author{Jana Giceva}
\email{jana.giceva@in.tum.de}
\affiliation{%
    \institution{Technical University of Munich}
    \city{Munich}
    \country{Germany}
}

\author{Holger Fröning}
\email{holger.froening@ziti.uni-heidelberg.de}
\affiliation{%
    \institution{Heidelberg University}
    \city{Heidelberg}
    \country{Germany}
}

\begin{abstract}
Efficiently finding subgraph embeddings in large graphs is crucial for many application areas like biology and social network analysis.
Set intersections are the predominant and most challenging aspect of current join-based subgraph query processing systems for CPUs.
Previous work has shown the viability of utilizing FPGAs for acceleration of graph and join processing. 

In this work, we propose GraphMatch, the first genearl-purpose stand-alone subgraph query processing accelerator based on worst-case optimal joins (WCOJ) that is fully designed for modern, field programmable gate array (FPGA) hardware.
For efficient processing of various graph data sets and query graph patterns, it leverages a novel set intersection approach, called AllCompare, tailor-made for FPGAs.
We show that this set intersection approach efficiently solves multi-set intersections in subgraph query processing, superior to CPU-based approaches.
Overall, GraphMatch achieves a speedup of over $2.68\times$ and $5.16\times$, compared to the state-of-the-art systems GraphFlow and RapidMatch, respectively.
\end{abstract}


\maketitle

\section{Introduction}
\label{sec:introduction}
Subgraph query processing, used \eg for graph pattern matching, is an important workload in many application areas \cite{journals/vldb/SahuMSLO20}, like social network analysis \cite{journals/sociometh/SnijdersPRH06} and protein interaction network analysis \cite{journals/bioinformatics/PrzuljCJ06}, where all embeddings identical to a given query graph are found in a data graph.

\begin{figure}[bt]
	\centering
	\includegraphics[width=\linewidth]{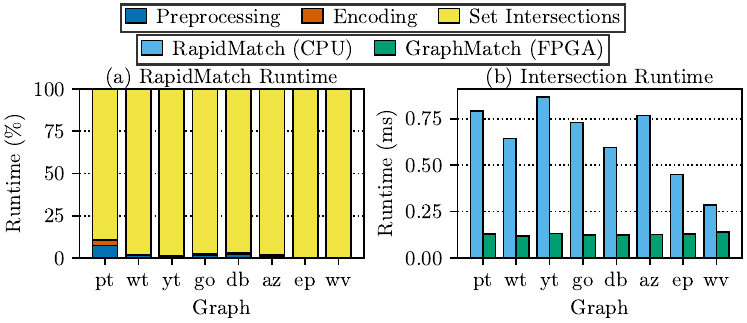}
	\caption{RapidMatch (CPU) runtime (left), and RapidMatch and GraphMatch (FPGA) intersection operators (right).}
	\label{fig:appetizer}
\end{figure}

Subgraph query processing was mainly approached with two kinds of algorithms in related work on CPUs, namely backtracking \cite{journals/jacm/Ullmann76, conf/sigmod/BiCLQZ16} and join-based approaches \cite{journals/tods/AbergerLTNOR17, journals/pvldb/MhedhbiS19}.
A recent study by Sun et al. \cite{conf/sigmod/Sun020} has shown that backtracking is efficient for large and sparse query graphs, and join approaches for smaller, dense query graphs.
For instance, one of the most advanced CPU-based systems, RapidMatch \cite{journals/pvldb/SunSC0H20}, is mainly based on joins, while considering graph structural information as in backtracking.
When looking into join-based approaches like RapidMatch, set intersections are the most expensive operation, as shown in \cref{fig:appetizer}(a).
Despite vectorized processing, this is due to challenges like limited parallelism, expensive round trips to main memory, and cache pollution in current general-purpose CPU hardware.

Specialized hardware like field programmable gate arrays (FPGAs) showed that they can solve these challenges with massive, unstructured data and pipeline parallelism and flexible data movement through configurable data flow architectures (\eg in related domains like graph processing \cite{conf/fpl/Dann0F22}, JSON parsing \cite{conf/damon/DannW0FF22}, and relational join processing \cite{conf/edbt/LaschMMDFS22}).
For instance, for subgraph matching, data movement could be more efficient by keeping partial matchings mostly on the FPGA chip.
The benefits of such hardware are shown in \cref{fig:appetizer}(b), which denotes a comparison of RapidMatch using SIMD-based set intersection (cf. \cite{conf/sigmod/Han0Y18}) and an FPGA-based implementation that will be explained in more detail subsequently.
However, in a recent survey on non-relational data processing, Dann et al. \cite{journals/csur/DannRF23} identified a gap for subgraph query processing on FPGAs, which was also acknowledged and partially addressed by Jin et al. \cite{conf/icde/JinY0YQP21} on subgraph query processing on a hybrid CPU-FPGA system.

Recent work on worst-case optimal join (WCOJ)-based subgraph query processing on CPUs \cite{journals/pvldb/MhedhbiS19} was very promising.
Thus, in this work, we focus on join-based subgraph query processing on FPGAs using WCOJs.
We propose GraphMatch, a graph processing accelerator for subgraph query processing fully implemented on an FPGA.
In GraphMatch, we first conceptually adapt the widely-used LeapFrog algorithm \cite{journals/corr/abs-1210-0481} to FPGAs, before specifying a novel, FPGA-native AllCompare algorithm, which leverages the FPGA's massive pipelining to its extreme.
GraphMatch is designed for (i) dense, small query graphs, (ii) supporting subgraph homomorphism and isomorphism, (iii) parallel, efficient set intersections, and (iv) directed and undirected graphs.
For that, GraphMatch implements a set of intersection operators with configurable number of input sets which are connected with partial matching multiplexers and demultiplexers able to dynamically switch matchings to a memory sink.
The GraphMatch query parser can generate query plans for arbitrary subgraph queries on-the-fly.
With these query plans we select how we chain together and execute the operators.
With GraphMatch, whose FPGA and subgraph query processing foundations and related work are introduced in \cref{sec:background}, we make the following contributions (\emph{Cx}):
\begin{itemize}
    \item[C1] We specify and implement two novel intersection accelerators tailored to FPGAs (LeapFrog and AllCompare). (\cref{sec:intersection})
    \item[C2] We design a subgraph query processing accelerator based on worst-case optimal joins, called GraphMatch. (\cref{sec:instance})
    \item[C3] We make GraphMatch configurable for dynamic queries and propose three performance optimizations. (\cref{sec:optimizations})
\end{itemize}

The resulting GraphMatch system shows promising scalability with an average speedup of $2.68\times$ over GraphFlow and $5.16\times$ over RapidMatch with a maximum speedup of over $100\times$ (\cref{sec:evaluation}).
Overall, we conjecture that FPGAs are well suited to solve the set intersection bottleneck of CPU-based subgraph query processing systems.
However, we still see areas of improvements for instance work balancing and for highly degree-skewed graphs (\cref{sec:conclusion}).

\section{Background and Related Work}
\label{sec:background}
In this section, we briefly introduce FPGAs and fundamentals of subgraph query processing, and discuss GraphMatch in the context of related work.

\subsection{Field Programmable Gate Arrays}
\label{sec:fpga}
\begin{figure}[bt]
	\centering
	\includegraphics[width=.85\linewidth]{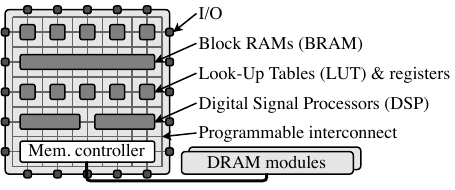}
	\caption{FPGA architecture (taken from \cite{conf/damon/DannW0FF22}).}
	\label{fig:fpga}
\end{figure}
Field programmable gate arrays (short FPGAs) map custom digital circuit designs (a set of logic gates and their connections) to a grid of resources (\ie look-up tables, registers) connected with a programmable interconnection network (cf. \cref{fig:fpga}).
For frequently used complex functionality like floating point computation, FPGAs contain digital signal processors.
Access to off-chip resources like DRAM and network controllers is possible over I/O pins.
The memory hierarchy of FPGAs is split into on-chip and off-chip memory.
On-chip, FPGAs implement distributed memory, which is made up of (a) single registers mostly used as storage for working values, and (b) block RAM (BRAM) in the form of SRAM memory components, mostly used for fast storage of data structures.
On modern FPGAs, there is about as much BRAM as cumulative cache on CPUs (all cache levels combined), but contrary to the fixed cache hierarchies of CPUs, BRAM is finely configurable to the needs of a given application.

\subsection{Subgraph Query Processing}
Graphs are abstract data structures ($G = (V,E)$) comprising of a vertex set $V$, and an edge set $E \subseteq V \times V$. The edges of directed graphs are denoted by a set $E$ of tuples.

\begin{figure}
    \centering
    \includegraphics[width=.8\linewidth]{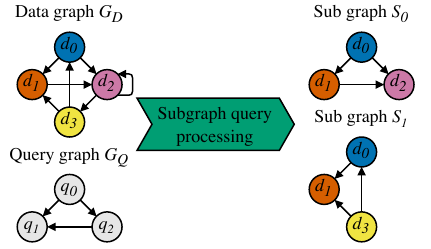}
    \caption{Subgraph query processing example and all its isomorphisms.}
    \label{fig:PatternMatching_ex}
\end{figure}

\Cref{fig:PatternMatching_ex} shows an example of an unlabeled subgraph query processing task for directed graphs, given a data graph $G_{D} = (V_{D},E_{D})$ with four vertices and seven edges, and an input query graph $G_{Q} = (V_{Q},E_{Q})$ with three vertices and three edges:
\begin{align*}
    V_{D} = &\{d_0, d_1, d_2, d_3\}\\
    E_{D} = &\{(d_0,d_1), (d_1,d_2), (d_2,d_3), (d_2,d_2), (d_3,d_0), (d_0,d_2), (d_3,d_1)\}\\
    V_{Q} = &\{q_0, q_1, q_2\}\\
    E_{Q} = &\{(q_0,q_1), (q_0,q_2), (q_2,q_1)\} \enspace .\\
\end{align*}
The task of subgraph query processing needs to compute either the homomorphisms or isomorphisms of the query graph within the data graph \cite{journals/pvldb/SunSC0H20}. Both denote subgraphs of the same shape as the query graph, but homomorphisms allow duplicate vertices within subgraphs, while isomorphisms do not \cite{journals/pvldb/SunSC0H20}.

In the given example, we identify all ismorphisms of the triangular $G_{Q}$ within $G_{D}$. 
Thus, all results of the task are triangular subgraphs of $G_{D}$, where edges of the same direction exist in the data graph and no vertex is used multiple times. 
The result contains two graphs:
\begin{align*}
    S_0 = (V_{S_0},E_{S_0}) := &(\{d_0,d_2,d_1\},\{(d_0,d_2),(d_0,d_1),(d_1,d_2)\})\\
    S_1 = (V_{S_1},E_{S_1}) := &(\{d_3,d_1,d_0\},\{(d_3,d_1),(d_3,d_0),(d_0,d_1)\}) \enspace .
\end{align*}
The homomorphisms in the example \cref{fig:PatternMatching_ex} include $S_0$ and $S_1$ and four subgraphs with multiple occurrences of vertex $d_2$:
\begin{align*}
    S_2 = (V_{S_2},E_{S_2}) := &(\{d_0,d_2\},\{(d_0,d_2),(d_2,d_2)\})\\
    S_3 = (V_{S_3},E_{S_3}) := &(\{d_1,d_2\},\{(d_1,d_2),(d_2,d_2)\})\\
    S_4 = (V_{S_4},E_{S_4}) := &(\{d_2\},\{(d_2,d_2)\})\\
    S_5 = (V_{S_5},E_{S_5}) := &(\{d_2,d_3\},\{(d_2,d_3),(d_2,d_2)\} ) \enspace .
\end{align*}
All results consist of vertices and edges in $G_{D}$ and have the same shape as $G_{Q}$. 
In this example, we colored each vertex of the data graph $G_{D}$ differently, for easier validation of the results. 
In general, all subgraphs must guarantee $\forall i: V_{S_i} \subseteq V_{D}$ and $\forall i: E_{S_i} \subseteq E_{D}$.

\paragraph{Exploration-Based Subgraph Query Processing}
Exploration-based algorithms (also known as backtracking) are one of the two main approaches used to process subgraph queries. 
The general idea is to explore the entire graph and create candidate sets of appropriate data vertices for each query vertex.
The next step takes those sets and enumerates all valid isomorphisms.

In the literature, three different enumeration variations are known \cite{journals/pvldb/SunSC0H20}: direct-, index-, and preprocessing-enumeration.
Their underlying ideas are the same, but they avoid or handle the start of the algorithm differently (\eg enumerate subgraphs directly, create an index structure on the data graph beforehand).
The general backtracking algorithm works similar for all variations. 
A candidate set along a query vertex ordering (QVO) is computed containing all data vertices that might be a valid entry for the given position \cite{journals/pvldb/SunSC0H20}.
Additional information, like edge connections between candidates, is also collected in separate data structures \cite{journals/pvldb/SunSC0H20}. Afterwards, the algorithm uses the collected information and data structures to enumerate all subgraph isomorphisms along the query vertex order recursively \cite{journals/pvldb/SunSC0H20}. 
Each iteration computes a local candidate set by taking the connections between previous and future vertices into account \cite{journals/pvldb/SunSC0H20}.
The recursion ends after the whole query graph is processed.
The algorithm can also be adapted to find homomorphisms instead by allowing duplicate vertices during the enumeration \cite{journals/pvldb/SunSC0H20}.

\paragraph{Join-Based Subgraph Query Processing}
Worst-case optimal joins (WCOJ) limit their running time to the worst-case output size of the algorithm \cite{journals/jacm/NgoPRR18,journals/pvldb/SunSC0H20,journals/pvldb/FreitagBSKN20}.
Join-based subgraph query processing is based on the WCOJ algorithm Generic Join \cite{journals/jacm/NgoPRR18, conf/sigmod/FuchsBG20, journals/pvldb/SunSC0H20,journals/pvldb/MhedhbiS19, journals/sigmod/NgoRR13}.
Generic Join describes an iterative approach to construct all homomorphisms by joining one vertex at a time to a temporary subgraph (partial matching). 

A known variation of Generic Join is Leapfrog Triejoin \cite{journals/corr/abs-1210-0481}. 
Leapfrog starts with a simple subgraph and extends it step by step.
To find valid join candidates, it intersects the corresponding neighbourhoods of all previously matched vertices that share an edge with the new vertex in the query graph. 
All elements of the result set are joined to the current subgraph. 
This process creates multiple new graphs for the next iteration with the trie structure of the algorithm. 
After each of the subgraphs represents a valid matching for the full query graph or no subgraph can be extended anymore, the algorithm terminates.
The intersections of Leapforg Triejoin are computed by its Leapfrog Join algorithm. 
This algorithm searches for potential result values in ordered sets. 
It leaps from one value to the other within sets and jumps from set to set to exclude values that can not be part of the intersection result.

To compute isomorphisms with a join-based approach, an additional check during the join phase is required \cite{journals/pvldb/SunSC0H20}. This paper focuses on identifying isomorphisms with a Join-based approach on FPGA, with simple reconfiguration to homomorphisms for comparison to the related GraphFlow \cite{journals/pvldb/MhedhbiS19} system.

\subsection{Related Work}
\label{sec:related-work}
\begin{table*}[ht]
\caption{Subgraph query processing systems, design principles, and properties.}
\label{tab:systems}
\small
\centering
\begin{tabular}{l l l l c c c c c}
 Name & Platform & Approach & Key data structure & General & Parallel & Dir. & Hom. & Iso. \\
 \hline
 \hline
 CFLMatch \cite{conf/sigmod/BiCLQZ16} & CPU & Backtracking & Compact path index & \faThumbsUp & \faThumbsODown & \faThumbsODown & \faThumbsODown & \faThumbsUp \\
 DAF \cite{conf/sigmod/HanKGPH19} & CPU & Backtracking & Candidate space & \faThumbsUp & \faThumbsUp & \faThumbsODown & \faThumbsODown & \faThumbsUp \\
 FAST \cite{conf/icde/JinY0YQP21} & FPGA \& CPU & Backtracking & Candidate search tree & \faThumbsUp & \faThumbsUp & \faThumbsODown & \faThumbsODown & \faThumbsUp \\
 \hline
 GraphZero \cite{journals/sigops/MawhirterRHLW21} & CPU & Nested loops & Adjacency lists & \faThumbsODown & \faThumbsODown & \faThumbsUp & \faThumbsODown & \faThumbsUp \\
 \hline
 EmptyHeaded \cite{journals/tods/AbergerLTNOR17} & CPU & WCOJ & Trie & \faThumbsODown & \faThumbsUp & \faThumbsUp & \faThumbsUp & \faThumbsODown \\
 GraphFlow \cite{journals/pvldb/MhedhbiS19} & CPU & WCOJ \& binary joins & Adjacency lists & \faThumbsUp & \faThumbsUp & \faThumbsUp & \faThumbsUp & \faThumbsODown \\
 CECI \cite{conf/sigmod/BhattaraiLH19} & CPU & Intersections & Compact embedding cluster index & \faThumbsUp & \faThumbsUp & \faThumbsUp & \faThumbsODown & \faThumbsUp \\
 RapidMatch \cite{journals/pvldb/SunSC0H20} & CPU & WCOJ \& hash joins & Encoded trie & \faThumbsUp & \faThumbsODown & \faThumbsODown & \faThumbsUp & \faThumbsUp \\
 \hline
 \textbf{GraphMatch} & FPGA & WCOJ & Compressed sparse row & \faThumbsUp & \faThumbsUp & \faThumbsUp & \faThumbsUp & \faThumbsUp \\
\end{tabular}

\medskip
Dir.: Directed graphs, Hom.: Subgraph homomorphisms, Iso.: Subgraph isomorphisms, \faThumbsUp: yes, \faThumbsODown: no

\end{table*}
Computing subgraph isomorphisms and homomorphisms is an important task studied in related work. 
Surveys, like Sun et al. \cite{conf/sigmod/Sun020}, explored different methods, approaches, and optimizations of subgraph matching. 
Lee at al. \cite{journals/pvldb/LeeHKL12} introduced generalized models and techniques for the general case of subgraph queries. 
The result of the increased attention in research for subgraph query processing is a variety of approaches and systems. 
\Cref{tab:systems} depicts an overview of the current state-of-the-art of subgraph query processing systems.

For backtracking-based systems, CFLMatch \cite{conf/sigmod/BiCLQZ16} and DAF \cite{conf/sigmod/HanKGPH19} are instrumental and use their custom candidate data structures to allow subgraph matching of general query graphs.
CFLMatch decomposes the query graph into a core, a forest and leaves that are both matched in that order because of their decreasing selectivity. 
DAF uses a candidate space data structure with dynamic programming, an adaptive query vertex ordering, and failing set pruning.
The system FAST \cite{conf/icde/JinY0YQP21} introduces a graph processing system for FPGA. 
It computes its candidate search tree data structure on CPU prior to FPGA computation \cite{conf/icde/JinY0YQP21}.
After the transfer of the structure to the FPGA's BRAM, the FPGA enumerates all subgraphs of the data graph for the given query \cite{conf/icde/JinY0YQP21}. 
Additionally, it allows concurrent computation with the host CPU to further speedup the computation \cite{conf/icde/JinY0YQP21}. 
GraphZero \cite{journals/sigops/MawhirterRHLW21} is a compilation based approach that tries to eliminate redundant computations in a nested loop structure.

EmptyHeaded \cite{journals/tods/AbergerLTNOR17} introduced the WCOJ approach to subgraph query processing systems as a compilation-based system and uses its trie data structure to support subgraph queries on directed graphs.
Additionally, EmptyHeaded supports graph processing.
GraphFlow \cite{journals/pvldb/MhedhbiS19} extends EmptyHeaded's approach by combining it with binary joins into a hybrid approach with a query optimizer.
This allows query plans which combine multiple partial embeddings with the binary join.
CECI \cite{conf/sigmod/BhattaraiLH19} splits up the data graph into embedding clusters to parallelize execution and emplys pruning to reduce the number of intermediate matchings.
RapidMatch \cite{journals/pvldb/SunSC0H20} is the most recent subgraph query processing system that proves that the backtracking and WCOJ approaches are complexity-wise equal.
Based on this observation, it combines a join-based approach with backtracking-like candidate pruning.

We propose GraphMatch, a pure FPGA design utilizing the WCOJ approach which can compute subgraph isomorphism and homomorphisms on directed or undirected graphs in parallel.

\section{Intersections on FPGAs}
\label{sec:intersection}
In this section we focus on designing a parallel, efficient set intersector for FPGAs motivated by the observation that set intersections are the most expensive operation of subgraph query processing (cf. \cref{fig:appetizer}(a)).
We describe the different set intersection approaches for CPUs and FPGAs on a spectrum from a software engineer's perspective to a hardware engineer's perspective (contribution \emph{C1}).
Thereafter, we introduce a highly FPGA-optimized implementation of AllCompare set intersection.
Lastly, we show how the set intersection approaches compare and characterize the performance dimensions of the AllCompare set intersector.

\subsection{Intersector Approaches for FPGAs}
\begin{figure}
    \centering
    \includegraphics[width=\linewidth]{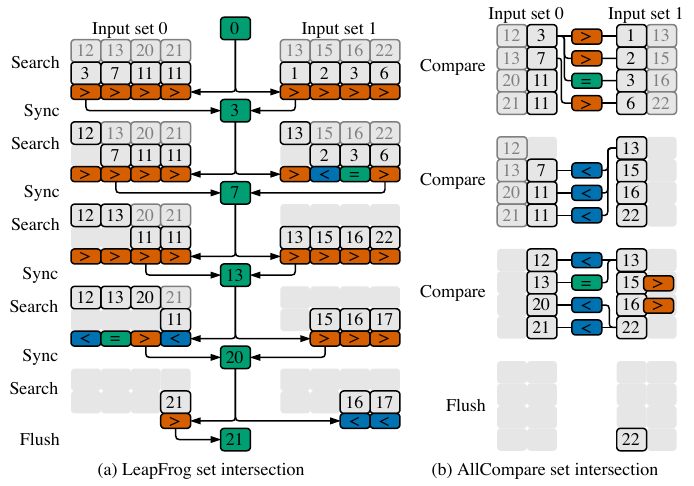}
    \caption{LeapFrog and AllCompare intersection approaches.}
    \label{fig:intersection_approaches}
\end{figure}
\Cref{fig:intersection_approaches} compares LeapFrog as the dominant set intersection approach for CPUs to the novel AllCompare set intersection approach specifically designed for FPGAs.

\paragraph{LeapFrog Set Intersection}
LeapFrog processes set intersections in turns of searching for a new search item and syncing the search item (\cref{fig:intersection_approaches}(a) shows an example).
The execution starts with $0$ as the search item (green box in the middle).
In each search step, the current search item is compared against each element in each input set (two in this example).
For the sync step, the next biggest element of each input set is communicated and compared to form the next search item and all elemets that are smaller than the previous search item are discarded.
This process is repeated until the search item is bigger than all remaining elements of on of the input sets.
The rest of the set elements are then flushed.
In each search and sync loop, LeapFrog has a guaranteed progress of only one element per set while the actual progress may be higher for real world sets.
We implement a parallelized version of LeapFrog set intersection in OneAPI that is able to perform all comparisons in the search step in one clock cycle and implement a VHDL version that does the same parallelization and additionally does input set prefetching which is not easily implementable in OneAPI.

\paragraph{AllCompare Set Intersection}
With the observation in mind, that on an FPGA we can implement many comparison operators in parallel, we propose the novel AllCompare set intersection approach for FPGAs.
\cref{fig:intersection_approaches}(b) shows how AllCompare is able to massively reduce the number of steps for the same set intersection that is shown for LeapFrog.
In each compare step, AllCompare compares all elements of both input sets against each other.
Elements that are smaller than an element in the other set are discarded, elements that have an equal element in the other sets are put out and discarded and all other elements remain.
In the end, the last remaining elements are flushed.
AllCompare guarantees progress of at least one full line of one of the input sets and is thus able to process the same intersection in under half of the clock cycles in this example.

\subsection{AllCompare Intersector Architecture}
\begin{figure}
    \centering
    \includegraphics[width=.75\linewidth]{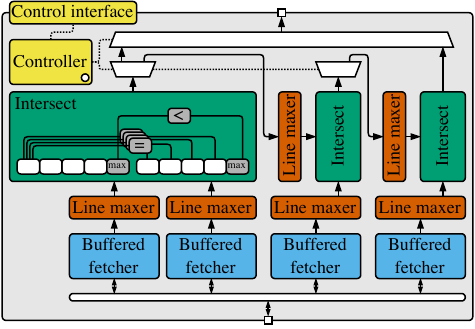}
    \caption{AllCompare set intersector architecture.}
    \label{fig:allcompare_architecture}
\end{figure}
\cref{fig:allcompare_architecture} shows the AllCompare set intersector architecture for four input sets.
Four buffered fetchers read the input data into the intersector and feed the input lines ($16$ elements per line on the FPGA $4$ elements per line in this example) through line maxers.
The line maxers find the maximum of the line which may not be the last element because not all elements of a line have to be valid (\eg set is smaller than line width).
These lines with extracted maximums are fed into the intersect operators.
In each intersect operator, in each cycle, all elements are equal compared against all elements from the other input set to determine which elements should be put out as the result of the set intersection.
Additionally, the maximums of both lines are discarded and the line with the smaller maximum is completeley discarded as all elements have to be smaller than at least one of the other line.
The results are fed into a demultiplexer which either forwards the output to the next line maxer and intersect operator or the output port.
As the last component, each AllCompare set intersector contains a controller that is connected to the control interface.
The user may define the switches that switch the demultiplexers and the multiplexer before the execution.
This allows dynamic reconfiguration of number of input sets during runtime.

\begin{figure}
    \centering
    \includegraphics[width=.5\linewidth]{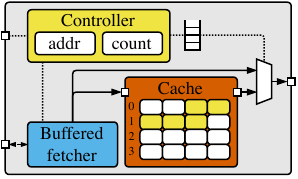}
    \caption{Cached fetcher architecture.}
    \label{fig:cached_fetcher}
\end{figure}
Especially during subgraph query processing, oftentimes the same vertex neighborhoods are accessed as input sets of the set intersector repeatedly.
Thus, we propose a cached fetcher architecture that stores the most recently accessed input set in a cache and serves subsequent requests to the same input set from the cache.
\cref{fig:cached_fetcher} shows the cached fetcher architecture.
The controller receives the requests and stores the last address and number of elements accessed in registers. 
If the request is equal to the previous one, it is flagged as cached and directly inserted into a FIFO queue which multiplexes the output port.
If the request cannot be served through the cache, it is forwarded to a buffered fetcher which send the request to memory and flagged as fetched and also inserted into the FIFO queue.
When the request is served by memory, the data is written into the cache (implemented as BRAM) as whole lines of memory starting from cache address $0$.
The FIFO queue is continuously observed. 
If a new request arrives, the flag if it is cached or not is read out and either the data from the buffered fetcher is directly forwarded or the respective number of memory lines are read from the cache again starting at address $0$.
The cached fetcher has the same interface as the buffered fetcher and may thus directly replace those in the AllCompare set intersector architecture.

\subsection{Intersector Performance Characteristics}
\label{sub:intersector_characteristics}
Finally, we discuss different performance characteristics of set intersection on FPGAs starting with a comparison of the RapidMatch intersection function, LeapFrog implemented in OneAPI, LeapFrog implemented in VHDL, and the AllCompare intersector.
Thereafter, we discuss the performance of AllCompare in detail for characteristics such as input set size, output set size, and number of input sets and degree of caching with the cached fetcher.

\paragraph{The benefits of adapting algorithms to FPGAs}
\begin{figure}[bt]
	\centering
	\includegraphics[width=\linewidth]{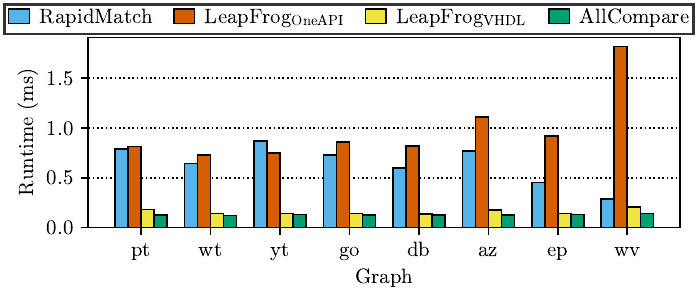}
	\caption{CPU (RapidMatch) vs. FPGA intersection operators.}
	\label{fig:inter_comp}
\end{figure}
\Cref{fig:inter_comp} shows the comparison of the CPU-based RapidMatch intersection function against the different FPGA-based implementations of intersectors that we introduced.

The benchmark environment as well as the graphs (cf. \cref{tab:graphs}) used are described in \cref{sec:setup}.
For LeapFrog$_{\textrm{OneAPI}}$, an even newer Agilex FPGA is used which, however, was not made available by Intel for the VHDL-based design flow during the work on this paper.
The benchmark shows the runtime in milliseconds of $5000$ intersections of neighborhoods of random vertices of the respective graph.

Overall, we observe that LeapFrog$_{\textrm{OneAPI}}$ performs similar to the RapidMatch intersection function for some graphs and worse for others.
The RapidMatch intersection function performance benefits from very small graphs that fit mostly into the cache hierarchy of the CPU (\ie the epinions (ep) and wiki-vote (wv) graphs).
The performance of LeapFrog$_{\textrm{OneAPI}}$ is heavily influenced by average degree of the graph, thus, the performance for the amazon (az) and wiki-vote (wv) graphs are noticeably worse than for the other graphs.
Additionally, the VHDL FPGA implementations LeapFrog$_{\textrm{VHDL}}$ and AllCompare (\ie specifically tailored to FPGAs) perform significantly better than the CPU and OneAPI implementations.
AllCompare always outperforms LeapFrog$_{\textrm{VHDL}}$.
Thus, we can see the progression from adopting a CPU algorithm with LeapFrog$_{\textrm{OneAPI}}$ in a software engineer-friendly environment over a VHDL implementation of the CPU algorithm (LeapFrog$_{\textrm{VHDL}}$) to a highly optimized FPGA implementation in AllCompare.
In the remainder of this paper, we proceed with the tailor-made AllCompare set intersector.

\paragraph{Characteristics of the tailor-made FPGA intersector}
\begin{figure}
    \centering
    \includegraphics[width=\linewidth]{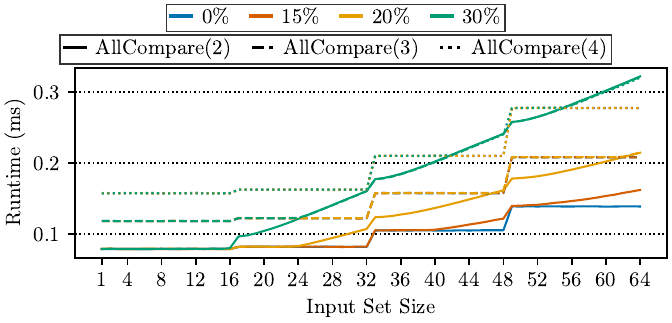}
    \caption{Runtime of set intersection with AllCompare for two to four input sets over input set size and output size as percentage of input size.} 
    \label{fig:inter_syn}
\end{figure}
\Cref{fig:inter_syn} shows the runtime of AllCompare in milliseconds with two, three, and four input sets on $5000$ set intersections with varying input set and output set size.
For AllCompare on two input sets, where $0\%$ of the input set are part of the output set (blue solid line) forms a memory-bound baseline.
The baseline is not influenced by fine-granular input set size but by the number of lines of memory it has to fetch.
For this implementation, $16$ elements of an input set fit into on memory line. 
Thus, the runtime increases each time the last element is part of a new memory line.
For the measurements where $15\%$, $20\%$, and $30\%$ of the input sets are in the output sets, the runtime is also bounded by the output set size because AllCompare only puts out one output set element per clock cycle.
More is not required by GraphMatch.
For three and four input sets, the runtime is higher and increasingly more bound by memory because more data has to be fetched from memory per set intersection.
For four input sets, only the measurement where $30\%$ of the input set are part of the output set (green dotted line) is sometimes bound by output size.

\begin{figure}
    \centering
    \includegraphics[width=.77\linewidth]{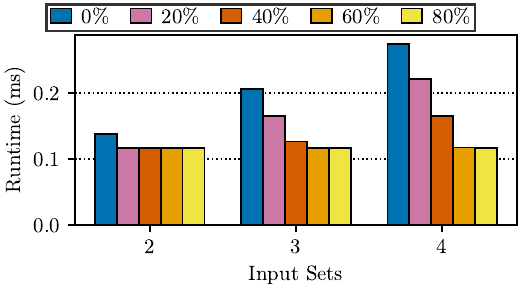}
    \caption{Input set caching for percentage of accesses cached by number of input sets.}
    \label{fig:cache_bench}
\end{figure}
\Cref{fig:cache_bench} shows how input set caching (cf. \cref{fig:cached_fetcher}) influences set intersection performance for the AllCompare set intersector for different number of input sets.
Again, we measure the runtime in milliseconds over $5000$ set intersections.
Each set intersection intersects sets of size $64$ without overlap such that output size does not influence performance.
Additionally, we vary cache hit rate from $0\%$ to $80\%$.
Overall, we observe that at the latest of $80\%$ cache hit rate, the performance reaches the same baseline for each number of input sets denoting a lower bound set by the cycles the FPGA logic needs to perform the intersection itself.
For larger number of input sets this baseline is reached with higher cache hit rate.
This is because more data has to be fetched from memory in the first place.
It is important to note that AllCompare has the same runtime irregardless of nubmer of input sets if memory requests do not play a role.

\subsection{Discussion}
This section provided a detailed look at the hardware design process in parts from the perspective of a software engineer to understand where FPGAs can provide benefits and that algorithms and data structures have to be specifically designed for the FPGA to provide good performance.
From the benchmarks, we conclude that the novel AllCompare set intersection approach provides the best performance and is thus used for our full subgraph query processing system GraphMatch.
The AllCompare set intersector performance is bound by memory which we optimize with a cached fetcher implementation that is able to provide increased performance for intersections with repeated input sets.

\section{GraphMatch}
\label{sec:graphmatch}
In this section, we first introduce the instance architecture of GraphMatch, our subgraph query processor, and its components (contribution \emph{C2}).
We then describe how subgraph queries can be dynamically switched in GraphMatch during runtime, and suitable performance optimizations that we applied to the base system (contribution \emph{C3}).

\subsection{GraphMatch Instance Architecture}
\label{sec:instance}
\begin{figure}
    \centering
    \includegraphics[width=\linewidth]{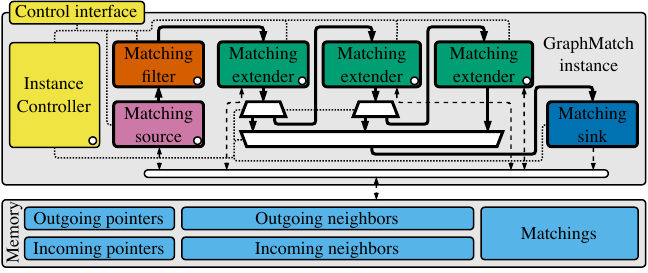}
    \caption{GraphMatch instance architecture.}
    \label{fig:graph_match}
\end{figure}
\Cref{fig:graph_match} depicts the architecture of a GraphMatch instance for subgraph matchings with up to five levels / vertices.
The flow of matchings through the architecture's components is depicted with bold arrows. 
It starts at the matching source -- producing matchings with two query vertices -- runs through a matching filter and multiple matching extenders (\ie a configuration with three extenders results in five levels overall), demultiplexers and multiplexers (cf. trapezoids) and ends at the matching sink.
The matching source reads all outgoing pointers and neighbors and combines them as edges, thereby forming the initial partial matchings.
Then the matching filter discards initial matchings that do not fit certain criteria like vertices not being distinct in the case of graph isomorphisms.
Each matching extender extends input partial matchings by one query vertex in order of the query vertex ordering, which most often means set intersections.
After each matching extender, there is a matching demultiplexer, which either forwards the incoming partial matchings to the next matching extender, or through a matching multiplexer to the matching sink.
Finally, the matching sink writes complete matchings to the designated matchings array in the FPGA's on-board memory.

The on-board memory contains in total five data arrays: 
one CSR data structure consisting of a pointers array and a neighbors array for both incoming and outgoing edges of each vertex, and the matchings array.
The matching source and matching extenders only read from memory, whereas the matching sink only writes to memory.
These accesses are shown in \cref{fig:graph_match} as dashed lines and are combined into one request stream fed to memory by a request merger (cf. white oval box).
The request merger also routes the memory read responses to the corresponding requesters.

The query graph is configured by providing GraphMatch with parameters.
All system components with white dots have parameters which are connected to the control interface operated by the CPU (shown by the dotted lines).
The matching source is parameterized with the \texttt{addresses} of the arrays it has to read.
The matching filter can be turned \texttt{on} and \texttt{off} with parameters and the matching extenders are also parameterized with memory \texttt{addresses} but also, for example, with the number of neighborhoods and which neighborhoods they should intersect.
Finally, the instance controller manages the query's execution.
It has a parameter for \texttt{query size} that switches the demultiplexers and multiplexer, is responsbile to trigger the execution when all components are ready, and finally returns relevant statistics to the CPU.

\begin{figure}
    \centering
    \includegraphics[width=\linewidth]{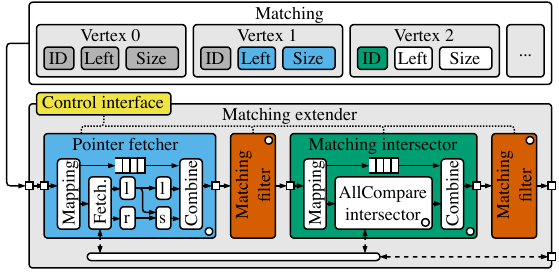}
    \caption{Matching data type and matching extender component.}
    \label{fig:matching_extender}
\end{figure}
\paragraph{Matching Data Type}
\Cref{fig:matching_extender} depicts the component design of a matching extender in the context of the matching data type.
Matchings each consist of a configurable number of vertices with a vertex identifier, left bound for pointers, and size that denotes the neighborhood size.
The number of vertices in the matching data type is equal to the number of levels in the GraphMatch instance (\eg five for the instance in \cref{fig:graph_match}).
The left bound and neighborhood size are kept in the matching as metadata between matching extenders.
Only when the query first requires intersection on the outbound edges of a vertex and then an intersection on the inbound edges or the other way around, do we have to fetch new metadata from the respective pointers array.
The metadata is discarded when writing to memory in the matching sink, since only the vertex identifiers are relevant for the result .

\paragraph{Matching Extender}
The matching extender component at level $l$ takes a matching with $l$ vertices where vertex at position $l$ is still missing the metadata.
It then fetches the required information and tries to extend the matching to vertex $l + 1$.
To do this, the partial matching is acquired through a pointer fetcher, retrieving required metadata, a first matching filter, a matching intersector performing the set intersection required and extending the matching by a vertex, and lastly another matching filter.
The pointer fetcher takes a partial matching and a \texttt{mapping} as a parameter and fetches the respective metadata.
Dictated by the mapping, a buffered fetcher (Fetch.) fetches the lines containing the pointers at positions $v$ and $v + 1$, where $v$ is the vertex identifier.
These form the left ($l$) and right ($r$) bound of the neighborhood.
The pointer fetcher subtracts $l$ from $r$ to get the neighborhood size and, finally, combines the new metadata with the partial matching.
The first matching filter filters out empty sets, \ie sets where any of the neighborhood sizes used in the following intersection are $0$.
The matching intersector maps the matching vertices to intersector spots in the AllCompare intersector, specified in \cref{sec:intersection}, based on \texttt{mapping} parameter extracted from the query graph.
For example, if there is an intersection between vertices $0$ and $2$, the AllCompare intersector is configured to do a $2$ set intersection mapping vertex $0$ to spot $0$ and vertex $1$ to spot $1$.
During the intersection, the partial matching is stored in a FIFO queue and the combined subcomponent extends the current partial matching until the intersection is finished.
It then proceeds with the next partial matching from the FIFO queue.
Depending on whether the workload is a graph isomorphism or homomorphism, the second matching filter sieves out matchings for which the newly added vertex is different from all vertices already part of the matching.

\subsection{Dynamic Queries \& Optimizations}
\label{sec:optimizations}
Based on the GraphMatch architecture, we specify dynamic queries with our query parser and transformations of a query graph into query parameters for GraphMatch.
We also explain our three key optimizations for GraphMatch: input set caching, failing set pruning, and stride mapping.

\begin{figure}
    \centering
    \includegraphics[width=\linewidth]{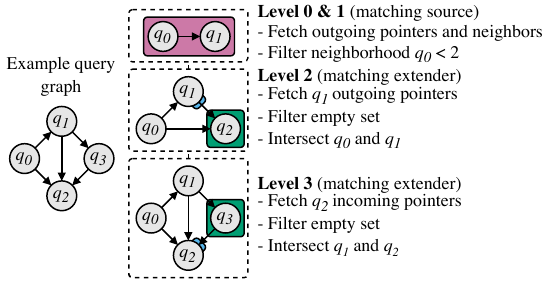}
    \caption{GraphMatch query parser for query graph Q5.}
    \label{fig:query_parser}
\end{figure}
\paragraph{Dynamic Queries}
\Cref{fig:query_parser} shows how the example query on the left side is deconstructed by the query parser to get the GraphMatch parameters to map the query graph to the system.
The instance controller receives the number of query vertices and address of the matchings array.
Each query starts with two vertices connected via an edge.
This forms levels $0$ and $1$ of the GraphMatch instance in the matching source which is parameterized with the addresses of the outgoing pointers and neighbors of $q_0$.
Additionally, the matching filter after the matching source receives the neighborhood size of $q_0$ in the complete query graph as a parameter.
On level $2$, the first matching extender is parameterized with a mapping to fetch the metadata for $q_1$ and the address to outgoing pointers for the pointer fetcher.
Additionally, the matching intersector receives a mapping to intersect the neighborhoods of $q_0$ and $q_1$ as a two-set intersection on outgoing neighbors.
Finally, for level $3$, the pointer fetcher should load the incoming pointers for $q_2$ and we pass a mapping to intersect the neighborhoods of $q_1$ and $q_2$ as a two set intersection on outgoing neighbors.
If a query is larger than the number of levels of the instance, we can materialize the partial matchings into memory, read them back to the beginning of the matching extender pipeline and feed them through the levels.
However, details of this are beyond the scope of this work.

\paragraph{Input set caching.} As a first optimization, we implement caching (cf. \cref{sub:intersector_characteristics}).
We suspect that locality of reference exists in the input set of the AllCompare intersector, as found in each matching intersector, as well as in the input of the pointer fetcher.
This is especially the case when new metadata for existing vertices has to be loaded for a partial matching.
Thus, all instances of the AllCompare intersector and the pointer fetcher employ caching.

\paragraph{Failing set pruning.}
As a second optimization, we introduce failing set pruning \cite{conf/sigmod/HanKGPH19}. In GraphMatch we implement it inside the matching filter, after the matching source, and in the first matching filter of each matching extender.
In addition to filtering out empty sets, we can also filter partial matchings when the neighborhood of a vertex is not at least as big as the corresponding vertices neighborhood in the query graph.
For example, for graph isomorphisms on Q5 (cf. \cref{fig:queries}), for $q_0$ the neighborhood size needs to be at least $2$.
Thus, we parametarize the matching filters for each vertex, so we can customize them for the query vertex neighborhood size.

\begin{figure}
    \centering
    \includegraphics[width=.8\linewidth]{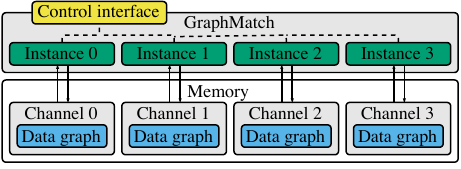}
    \caption{GraphMatch multi-instance scaling.}
    \label{fig:graphmatch_system}
\end{figure}
\paragraph{Parallelism.}
\Cref{fig:graphmatch_system} shows the GraphMatch system scaled to four instances.
Each instance is assigned its own memory channel and otherwise only connected to the control interface.
The data graph is copied to each memory channel in whole such that the instances can work truly independent.
The system can, thus, either process a query on one data graph in parallel such that the vertex set is split up into four intervals each assigned to the one of the four instances, or process different combinations of queries and data graphs on the four instances independently.

\paragraph{Stride mapping.}
As a last optimization, we apply stride mapping \cite{conf/fpga/DaiHCXWY17} to improve workload load balancing across instances of GraphMatch.  Load balancing was found to be crucial when processing complex datasets.
As the GraphMatch instances cannot communicate partial matching among each other, each vertex interval should require roughly equal amount of work.
Unfortunately, this is not a realistic assumption when working with real world graphs \cite{conf/fpl/Dann0F22}, which are often skewed. Here our optimization for stride mapping comes into play, by doing a light-weight vertex reordering.
Stride mapping is a technique for semi-random shuffling of the vertex identifiers before partitioning to create a new vertex ordering with a constant stride. In our case we use a stride of $100$, which results in a new vertex order $v_0, v_{100}, v_{200}, ...$, which virtually results in each vertex being mapped to a different hardware instance.

\section{Evaluation}
\label{sec:evaluation}
In this section, we first introduce the system used for the evaluation, the benchmark setup and key metrics such as resource utilization and clock frequency of the design, the graph data sets, and the graph queries.
We then report benchmark results scaling GraphMatch from $1$ up to $4$ instances, compare GraphMatch against the state-of-the-art CPU-based systems GraphFlow and RapidMatch, and evaluate the effects of different optimizations employed in GraphMatch on overall performance.

\subsection{Setup}
\label{sec:setup}
\begin{figure}[bt]
    \centering
    \includegraphics[width=\linewidth]{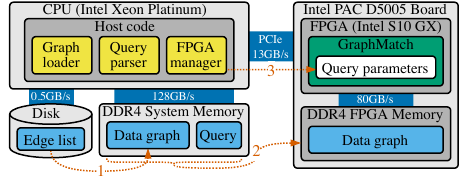}
    \caption{Overall system architecture (incl. host, device, memory).}
    \label{fig:computer-architecture}
\end{figure}
\paragraph{System Architecture}
\Cref{fig:computer-architecture} shows how GraphMatch is deployed in the system context.
In principle, the system features a CPU and an accelerator board -- connected via PCIe -- which hosts the FPGA, running GraphMatch itself, and memory, used as intermediate data storage for the data graph during subgraph query processing.
The CPU loads and prepares the data graph, parses and programs the query graph to GraphMatch and manages the execution on the FPGA.
To execute a particular workload with a particular graph, the GraphMatch framework is first synthesized with the query parameter registers, a fixed number of instances and maximum query size (\ie levels).
Afterwards, the synthesized design is programmed to the FPGA.

\paragraph{Execution Flow}
For the execution of a particular subgraph query on a particular graph data set, the edge list (or any other representation) of the graph is read from disk to the CPU and brought into two CSR data structures, one for outgoing edges and one for incoming edges of each vertex (cf. step (1)). 
During loading of the data graph, we transform the set of vertex identifiers to be dense (\ie excluding vertices that have degree $0$).
The data graph is then replicated to each channel of the FPGA memory (cf. step (2)).
Thereafter, the query graph is parsed and the respective parameter registers of GraphMatch are programmed via a control interface such that GraphMatch performs the subgraph query matching the query graph.
The host code also triggers the execution via the control interface (cf. step (3)).
After the execution finished, the results can be read back to CPU memory and used for further processing.
If desired, the data graph or parameter register values can be used multiple times in a row by loading new parameters or a new data graph respectively and again triggering the execution.

\paragraph{Hardware Details}
For our experiments, we work with a server equipped with an Intel FPGA Programmable Accelerator Card (PAC) D5005 attached via PCIe version 3.
The system features two Intel Xeon Gold 6142 CPUs at 2.6GHz and 384GB of DDR4-2666 memory, while the D5005 board is equipped with four channels of DDR4-2400 memory with a total capacity of 32GB and a resulting bandwidth of $76.8$GB/s.
The design itself is based on the Intel Open Programmable Execution Engine (OPAE) platform and is synthesized with Quartus version 19.4.

{\setlength{\tabcolsep}{0.475em}
\begin{table}[bt]
\caption{Resource utilization and clock frequency by graph problem and number of graph cores.}
\label{tab:utilization}
\small
\centering
\begin{tabular}{l r r c|r r r r r}
 System & $p$ & $l$ & $c$ & LUTs & Regs. & BRAM & DSPs & Clock freq. \\
 \hline
 \hline
 \multirow{2}{*}{GraphMatch} & $4$ & $6$ & \faThumbsDown & $48$\% & $26$\% & $21$\% & $0$\% & $187$MHz \\
  & $4$ & $6$ & \faThumbsOUp  & $54$\% & $33$\% & $34$\% & $0$\% & $191$MHz \\
\end{tabular}

\medskip
LUTs: Look-up tables; Regs.: Registers; BRAM: Block RAM; DSPs.: Digital signal processors; Clock freq.: Clock frequency

\end{table}}
\paragraph{Prototype Configurations}
\cref{tab:utilization} shows the two different system configurations used for the benchmarks.
We synthesized one system variant without input set caching ($c$) and one with input set caching for the pointer fetchers and set intersectors.
Both variants have $p=4$ instances of GraphMatch, one for each memory channel, with a maximum query graph size of $l=6$.
Note that, without further resource utilization optimization, up to 8 instances could be placed onto the chip -- derived from \cref{tab:utilization} -- however, which could not adequately leverage the available memory channels (cf. discussion on HBM in \cref{sub:discussion}).
Since the instances are not connected, the resource utilization scales linearly excluding the fixed resource utilization of the OPAE wrapper.
All types including pointers and vertex identifiers are $32$-bit unsigned integers.
Lastly, the depth of the reorder stage is set to $32$.
The resource utilization leaves room to scale to modern high-bandwidth memory or include more functionality, like graph processing \cite{conf/fpl/Dann0F22}, in the accelerator.

\begin{table}[bt]
\caption{Graphs used often by systems in \cref{tab:systems} (real-world graphs from \cite{LeskovecK14} and \cite{conf/aaai/RossiA15}).}
\label{tab:graphs}
\small
\centering
\begin{tabular}{l r r r r r}
 Name & $|V|$ & $|E|$ & $D_{avg}$ & \o & SCC \\
 \hline
 \hline
 patents (pt) & $3.8$M & $16.5$M & $4.34$ & $22$ & $1.00$ \\
 wiki-talk (wt) & $2.4$M & $5.0$M & $2.10$ & $11$ & $0.05$ \\
 youtube (yt) & $1.2$M & $3.0$M & $5.16$ & $20$ & $0.98$ \\
 google (go) & $875.7$K & $5.1$M & $5.82$ & $21$ & $0.50$ \\
 dblp (db) & $426.0$K & $1.0$M & $4.93$ & $21$ & $0.74$ \\
 amazon (az) & $403.3$K & $3.4$M & $8.43$ & $21$ & $0.98$ \\
 epinions (ep) & $75.9$K & $508.8$K & $6.70$ & $14$ & $0.43$ \\
 wiki-vote (wv) & $7,115$ & $103.7$K & $14.56$ & $7$ & $0.18$ \\
 \hline
 syn$_{n, d}$ & $n$ & $n \cdot d$ & $d$ & - & 1 \\
\end{tabular}

\medskip
SCC: Ratio of vertices in the largest strongly-connected component to $n$; \faThumbsOUp: yes, \faThumbsDown: no

\end{table}
\paragraph{Data and Query Graphs}
\cref{tab:graphs} show the graph data sets used to benchmark GraphMatch.
This selection represents the most important graphs considered by the other state-of-the-art subgraph query processing systems (cf. \cref{sec:related-work}).
We additionally show graph properties like size ($|V|$ and $|E|$), average degree ($D_{avg}$), and ratio of vertices in the largest strongly-connected component (SCC) that are useful to explain different performance effects observed in the benchmarks.
For the intersection benchmark, we generated different configurations of a synthetic graph syn$_{n, d}$ with another parameter for output size of the resulting intersection between two adjacent vertices.

\begin{figure}[bt]
    \centering
    \includegraphics[width=\linewidth]{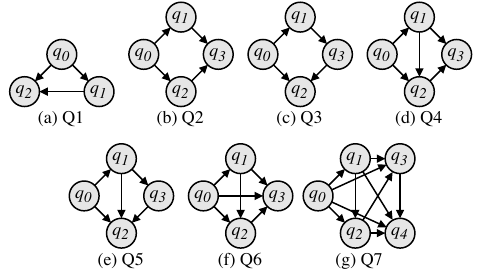}
    \caption{Query graphs (adopted from \cite{journals/pvldb/MhedhbiS19}).}
    \label{fig:queries}
\end{figure}
\cref{fig:queries} shows the query graphs we use in our evaluation, taken from \cite{journals/pvldb/MhedhbiS19}.
These can be classified as cliques (Q1, Q6, and Q7), cycles (Q1, Q2, and Q3), and other graphs (Q4 and Q5).

\subsection{GraphMatch Scalability}
\begin{figure*}[bt]
    \centering
    \includegraphics[width=\linewidth]{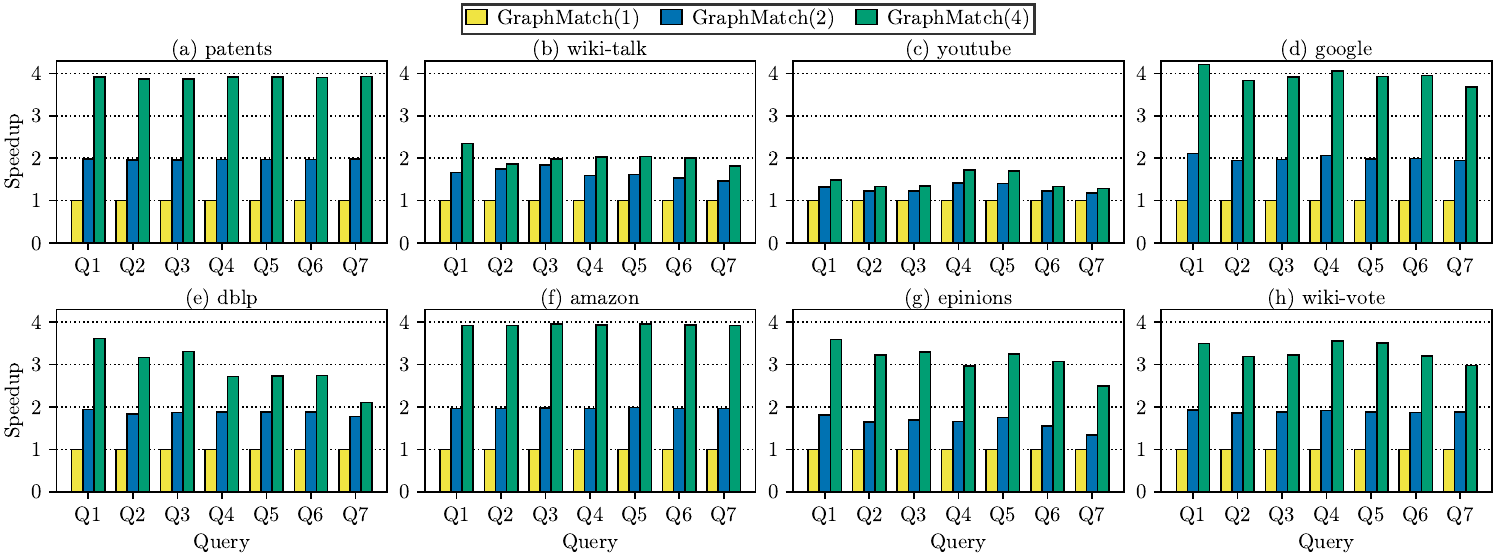}
    \caption{Scalability of GraphMatch from $1$ to $4$ instances.}
    \label{fig:scalability}
\end{figure*}
\cref{fig:scalability} shows the scalability of GraphMatch as we increase from a single up to four instances. We report the speedup over a baseline using a single instance for all data graphs and queries.
When using a single instance, the stride mapping optimization is disabled because it makes no difference for measurements on a single instance.
Otherwise, all optimizations discussed in \cref{sec:optimizations} are enabled by default.
The measurements show that the speedup is mostly dependent on the properties of the data set itself. 
For example, we observe a linear scalability  on the patents and amazon graphs, which is not the case for the other graphs, where scalability is influenced by the number of intermediate (and actual matchings) in the range of vertices assigned to an instance.
This effect is particularly pronounced for the wiki-talk and youtube graphs.
Scalability could be improved in future work, \eg with work-stealing where an idling instance takes over partial matchings from a busy instance to balance out the load.
However, this would also introduce dependencies between the instances and could lead to higher design complexity.

\subsection{Comparison to GraphFlow and RapidMatch}
We recall from \cref{sec:background} that the predominant CPU-based subgraph query processing systems are GraphFlow and RapidMatch, which we compare GraphMatch to subsequently.
\begin{figure*}[bt]
    \centering
    \includegraphics[width=\linewidth]{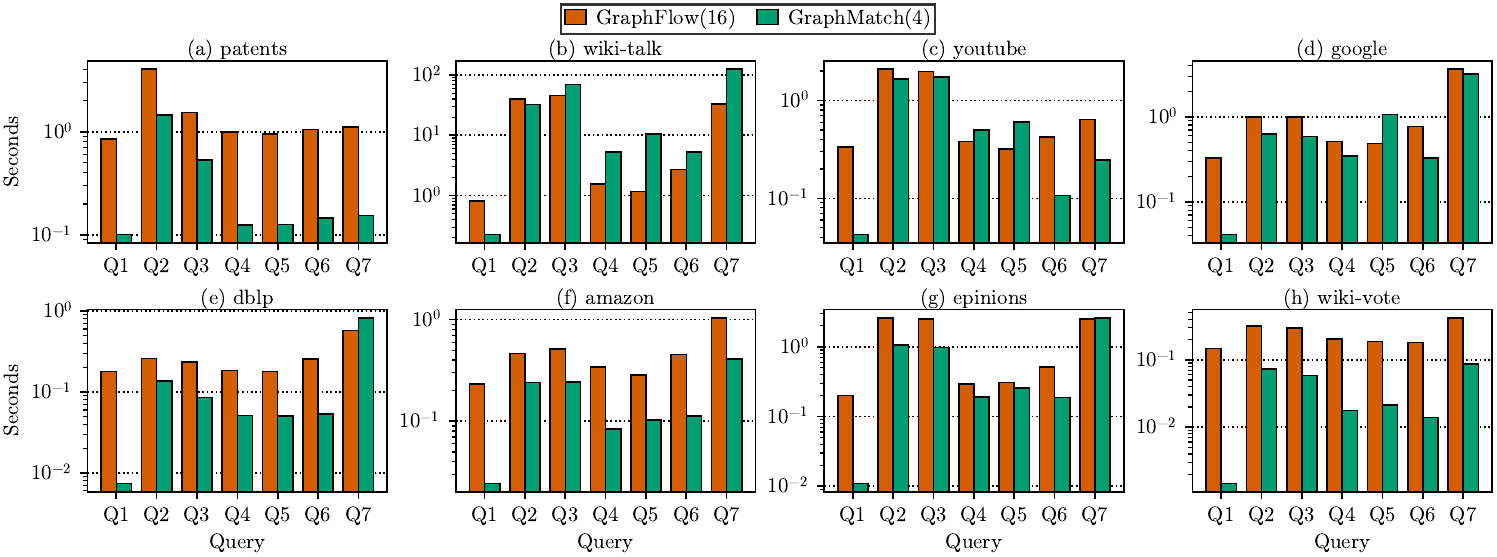}
    \caption{GraphMatch ($4$ instances) vs. GraphFlow ($16$ threads) on directed graphs computing subgraph homomorphisms.}
    \label{fig:comparison}
\end{figure*}
\paragraph{GraphFlow} \cref{fig:comparison} compares the performance of GraphMatch with four instances to GraphFlow \cite{journals/pvldb/MhedhbiS19} running on $16$ threads (a full CPU on our server) with  \texttt{numactrl} that restricts all threads to one NUMA node.
The plots show the performance in seconds of runtime on a logarithmic scale.
For GraphMatch, we tried out different QVOs for each query and data graph combination and show the best one.
This makes a huge difference and as future work can be added as a query optimizer step to the query parser.
GraphFlow also uses directed graphs, however, only supporting subgraph homomorphisms.
Thus, to set the ground for a fair comparison, we have turned off distinct vertex checking and changed the failing set pruning optimizations to match the workload in GraphMatch.
We note that for GraphFlow, we had to execute three scripts that prepare the graphs beforehand for each graph.
It was not clear to us how much these scripts optimize the graph layout.
When looking at the performance numbers, overall we observe big performance improvements with GraphMatch. 
This is especially pronounced for Q1 on all data graphs. 
GraphMatch also performs exceptionally well for Q4-Q6 on many graphs.
Another interesting observation is that Q2 and Q3 pose a challenge for GraphMatch on all graphs.
We attribute that to their low intermediate selectivity, which leads to a large number of partial matchings after extending to $q_2$.
Finally, we note that the graph structures of wiki-talk and youtube are more problematic for GraphMatch than for GraphFlow.
These graphs exhibit highly exponential (\ie skewed) degree distributions, which lead to some very large vertex neighborhoods which are used often in partial matchings.
These cannot be cached with our design which has relatively small caches and are able to be cached by the more sophisticated cache hierarchy of CPUs.
In particular, GraphMatch outperforms GraphFlow by an average of $2.68\times$ for all graphs with a maximum of $100\times$ for query-1 on wiki-vote. 

\begin{figure*}[bt]
    \centering
    \includegraphics[width=\linewidth]{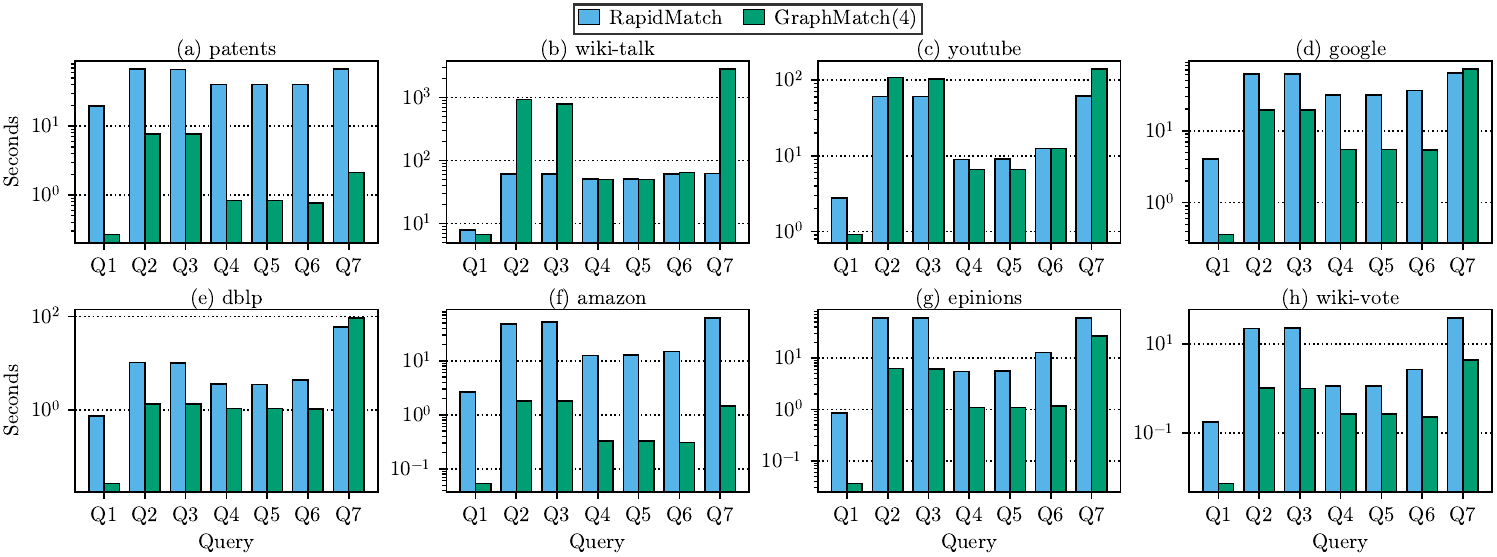}
    \caption{GraphMatch ($4$ instances) vs. RapidMatch on undirected graphs when computing subgraph isomorphisms.}
    \label{fig:rapidmatch}
\end{figure*}
\paragraph{RapidMatch.} \cref{fig:rapidmatch} compares GraphMatch ran with four instances to RapidMatch \cite{journals/pvldb/SunSC0H20}.
We were not able to find any configuration parameter to make RapidMatch run in parallel, so it only uses vectorized instructions for intra-intersection parallelism.
The plots show the runtime of both systems in seconds, and on a logarithmic scale.
RapidMatch only works with undirected graphs, so we also make the graphs undirected for GraphMatch and compute sub-graph isomorphisms (even though RapidMatch can also compute homomorphisms).
Overall, we observe significantly better performance for GraphMatch compared to RapidMatch, with an average speedup of $5.16\times$.
Similar to the comparison to GraphFlow, GraphMatch performs exceptionally well for Q1. As before, the system performs less well for 
Q2, Q3, and Q7. 
We further note that the performance results for Q2 and Q3, as well as for Q4 and Q5 are similar, because in an undirected graph the orientation of the query edges makes no difference.
Once again, we note the wiki-talk and youtube graphs are challenging for GraphMatch because of their heavy exponential degree distribution.

\subsection{Effects of GraphMatch Optimizations}
\begin{figure}[bt]
    \centering
    \includegraphics[width=\linewidth]{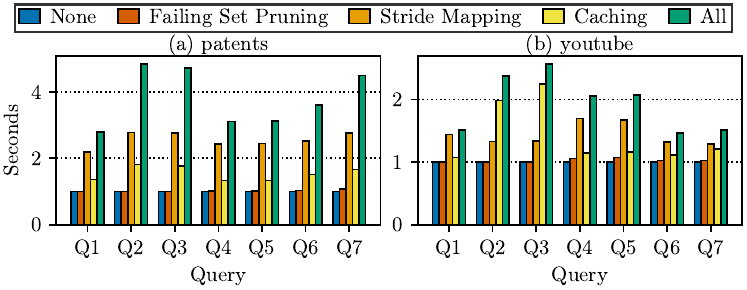}
    \caption{Effects of GraphMatch optimizations for patents and youtube data graphs.}
    \label{fig:optimizations}
\end{figure}
\cref{fig:optimizations} shows the effects of the failing set pruning, stride mapping, and input set caching optimizations on performance on the example of the patents and the youtube query graphs.
The baseline is GraphMatch with four instances with all optimizations turned off (None).
The failing set pruning optimization effectiveness depends on the query graph.
For most query graphs it has a small effect but gets more effective with larger query graphs (\eg Q7).
Stride mapping has the largest effect because it balances out the work required between the GraphMatch instances.
The effectiveness of input set caching mostly depends on the data graph.
It has the biggest effect on the youtube data graph benchmarks.
The optimizations work well together to provide a significant performance boost when all of them are turned on (All).

\subsection{Performance Model}
The effectiveness of input set caching showed that memory accesses are the bottleneck of GraphMatch.
Thus, we propose a performance model that is based on the number of memory requests with $l$ denoting the number of vertex identifiers or pointers that fit into the width of the memory interface. 
For instance, for $32$bit values and $512$bit wide memory interface, $l$ equals $16$.
To materialize the initial edges, we need the following number of memory requests:

\[(|V| + 1) / l + |E| / l\]

We need to read $|V| + 1$ pointers and $|E|$ edges sequentially.
Thus, we can divide these numbers by $l$.
For each extension of this partial matching we need approximately this additional amount of memory requests, where $f$ is the number of vertices for which new pointers have to be fetched, $m$ is the number of partial matchings going into this extension, and $s$ is the number of sets being intersected:

\[f \cdot m + s \cdot (m \cdot D_{avg} / \min (l, D_{avg}))\]

We need to fetch $f \cdot m$ pairs of pointers which however most of the time are part of one line of memory such that we only need one request.
For each intersector, we need to make $m \cdot D_{avg} / \min (l, D_{avg})$ requests.
The minimum term is because if the neighbor hood is smaller than $l$, we still have to fetch a whole line of memory.
However, this may be lowered with caching.
For example, $s=1$ for an extension by one edge,  and $m=|E|$ for the first extension after the initial edges.

\subsection{Discussion}

This section provided an in-depth analysis of the performance of GraphMatch on various graph datasets. 
We observe that the system exhibits linear scalability when the graphs have close to uniform degree distribution; and that the relative performance can be significantly improved with our proposed optimizations (up to 5x on the patents dataset and close to 3x on youtube). A detailed ablation study has demonstrated the effect of each individual optimization strategy and how their effects differ depending on a given query and input graph. When compared to state-of-the-art systems (GraphFlow and RapidMatch), we have shown that GraphMatch has superior performance, on average outperforming GraphFlow by $2.68\times$, and RapidMatch by $5.16\times$ across all queries and datasets. 
When zooming in on the performance for individual queries, we observe that GraphMatch exhibits an excellent performance for all queries when working with graph datasets that have an only slightly skewed degree distribution, and that there is room for improvement when handling cyclic queries like Q2 and Q3 on highly skewed graphs (e.g., youtube and wiki-talk).

\section{Conclusion}
\label{sec:conclusion}
We proposed GraphMatch, an FPGA-based efficient subgraph query processing accelerator.
GraphMatch is inspired by the insight that current CPU-based subgraph query processing systems are limited by set intersection which FPGAs are well suited for.
We showed the potential of such a design by first introducing a novel set intersector AllCompare specifically developed for FPGAs that is able to outperform all state-of-the-art CPU intersectors and FPGA adaptations of CPU set intersection algorithms.
Therafter, we introduce the GraphMatch architecture with dynamic query reconfiguration and three key performance optimizations.
Our experimental performance evaluation showed scalability and superior performance of GraphMatch compared to state-of-the-art CPU-based subgraph query processing systems GraphFlow and RapidMatch with an average speedup of $2.68\times$ and up to $5.16\times$ respectively.

In future work, we want to extend GraphMatch to benchmarks with labeled graphs and further improve performance for highly degree skewed graphs with more sophisticated caching.
Additionally, we consider connecting the GraphMatch instances with a work-stealing enabling matching crossbar and exploring query plan optimization for GraphMatch could unlock even higher performance.

\bibliographystyle{ACM-Reference-Format}
\bibliography{paper}


\begin{thebibliography}{27}


\ifx \showCODEN    \undefined \def \showCODEN     #1{\unskip}     \fi
\ifx \showDOI      \undefined \def \showDOI       #1{#1}\fi
\ifx \showISBNx    \undefined \def \showISBNx     #1{\unskip}     \fi
\ifx \showISBNxiii \undefined \def \showISBNxiii  #1{\unskip}     \fi
\ifx \showISSN     \undefined \def \showISSN      #1{\unskip}     \fi
\ifx \showLCCN     \undefined \def \showLCCN      #1{\unskip}     \fi
\ifx \shownote     \undefined \def \shownote      #1{#1}          \fi
\ifx \showarticletitle \undefined \def \showarticletitle #1{#1}   \fi
\ifx \showURL      \undefined \def \showURL       {\relax}        \fi
\providecommand\bibfield[2]{#2}
\providecommand\bibinfo[2]{#2}
\providecommand\natexlab[1]{#1}
\providecommand\showeprint[2][]{arXiv:#2}

\bibitem[Aberger et~al\mbox{.}(2017)]%
        {journals/tods/AbergerLTNOR17}
\bibfield{author}{\bibinfo{person}{Christopher~R. Aberger},
  \bibinfo{person}{Andrew Lamb}, \bibinfo{person}{Susan Tu},
  \bibinfo{person}{Andres N{\"{o}}tzli}, \bibinfo{person}{Kunle Olukotun},
  {and} \bibinfo{person}{Christopher R{\'{e}}}.}
  \bibinfo{year}{2017}\natexlab{}.
\newblock \showarticletitle{EmptyHeaded: {A} Relational Engine for Graph
  Processing}.
\newblock \bibinfo{journal}{\emph{{ACM} Trans. Database Syst.}}
  \bibinfo{volume}{42}, \bibinfo{number}{4} (\bibinfo{year}{2017}),
  \bibinfo{pages}{20:1--20:44}.
\newblock


\bibitem[Bhattarai et~al\mbox{.}(2019)]%
        {conf/sigmod/BhattaraiLH19}
\bibfield{author}{\bibinfo{person}{Bibek Bhattarai}, \bibinfo{person}{Hang
  Liu}, {and} \bibinfo{person}{H.~Howie Huang}.}
  \bibinfo{year}{2019}\natexlab{}.
\newblock \showarticletitle{{CECI:} Compact Embedding Cluster Index for
  Scalable Subgraph Matching}. In \bibinfo{booktitle}{\emph{{SIGMOD}}}.
  \bibinfo{publisher}{{ACM}}, \bibinfo{pages}{1447--1462}.
\newblock


\bibitem[Bi et~al\mbox{.}(2016)]%
        {conf/sigmod/BiCLQZ16}
\bibfield{author}{\bibinfo{person}{Fei Bi}, \bibinfo{person}{Lijun Chang},
  \bibinfo{person}{Xuemin Lin}, \bibinfo{person}{Lu Qin}, {and}
  \bibinfo{person}{Wenjie Zhang}.} \bibinfo{year}{2016}\natexlab{}.
\newblock \showarticletitle{Efficient Subgraph Matching by Postponing Cartesian
  Products}. In \bibinfo{booktitle}{\emph{{SIGMOD}}}.
  \bibinfo{publisher}{{ACM}}, \bibinfo{pages}{1199--1214}.
\newblock


\bibitem[Dai et~al\mbox{.}(2017)]%
        {conf/fpga/DaiHCXWY17}
\bibfield{author}{\bibinfo{person}{Guohao Dai}, \bibinfo{person}{Tianhao
  Huang}, \bibinfo{person}{Yuze Chi}, \bibinfo{person}{Ningyi Xu},
  \bibinfo{person}{Yu Wang}, {and} \bibinfo{person}{Huazhong Yang}.}
  \bibinfo{year}{2017}\natexlab{}.
\newblock \showarticletitle{{ForeGraph}: Exploring Large-scale Graph Processing
  on Multi-{FPGA} Architecture}. In \bibinfo{booktitle}{\emph{FPGA}}.
  \bibinfo{pages}{217--226}.
\newblock


\bibitem[Dann et~al\mbox{.}(2022a)]%
        {conf/fpl/Dann0F22}
\bibfield{author}{\bibinfo{person}{Jonas Dann}, \bibinfo{person}{Daniel
  Ritter}, {and} \bibinfo{person}{Holger Fr{\"{o}}ning}.}
  \bibinfo{year}{2022}\natexlab{a}.
\newblock \showarticletitle{{GraphScale}: Scalable Bandwidth-Efficient Graph
  Processing on {FPGAs}}. In \bibinfo{booktitle}{\emph{{FPL}}}.
  \bibinfo{publisher}{{IEEE}}, \bibinfo{pages}{24--32}.
\newblock


\bibitem[Dann et~al\mbox{.}(2023)]%
        {journals/csur/DannRF23}
\bibfield{author}{\bibinfo{person}{Jonas Dann}, \bibinfo{person}{Daniel
  Ritter}, {and} \bibinfo{person}{Holger Fr{\"{o}}ning}.}
  \bibinfo{year}{2023}\natexlab{}.
\newblock \showarticletitle{Non-relational Databases on FPGAs: Survey, Design
  Decisions, Challenges}.
\newblock \bibinfo{journal}{\emph{{ACM} Comput. Surv.}} \bibinfo{volume}{55},
  \bibinfo{number}{11} (\bibinfo{year}{2023}), \bibinfo{pages}{225:1--225:37}.
\newblock


\bibitem[Dann et~al\mbox{.}(2022b)]%
        {conf/damon/DannW0FF22}
\bibfield{author}{\bibinfo{person}{Jonas Dann}, \bibinfo{person}{Royden
  Wagner}, \bibinfo{person}{Daniel Ritter}, \bibinfo{person}{Christian
  Faerber}, {and} \bibinfo{person}{Holger Fr{\"{o}}ning}.}
  \bibinfo{year}{2022}\natexlab{b}.
\newblock \showarticletitle{{PipeJSON}: Parsing {JSON} at Line Speed on FPGAs}.
  In \bibinfo{booktitle}{\emph{DaMoN}}. \bibinfo{pages}{3:1--3:7}.
\newblock


\bibitem[Freitag et~al\mbox{.}(2020)]%
        {journals/pvldb/FreitagBSKN20}
\bibfield{author}{\bibinfo{person}{Michael~J. Freitag},
  \bibinfo{person}{Maximilian Bandle}, \bibinfo{person}{Tobias Schmidt},
  \bibinfo{person}{Alfons Kemper}, {and} \bibinfo{person}{Thomas Neumann}.}
  \bibinfo{year}{2020}\natexlab{}.
\newblock \showarticletitle{Adopting Worst-Case Optimal Joins in Relational
  Database Systems}.
\newblock \bibinfo{journal}{\emph{{PVLDB}}} \bibinfo{volume}{13},
  \bibinfo{number}{11} (\bibinfo{year}{2020}), \bibinfo{pages}{1891--1904}.
\newblock


\bibitem[Fuchs et~al\mbox{.}(2020)]%
        {conf/sigmod/FuchsBG20}
\bibfield{author}{\bibinfo{person}{Per Fuchs}, \bibinfo{person}{Peter~A.
  Boncz}, {and} \bibinfo{person}{Bogdan Ghit}.}
  \bibinfo{year}{2020}\natexlab{}.
\newblock \showarticletitle{EdgeFrame: Worst-Case Optimal Joins for
  Graph-Pattern Matching in Spark}. In
  \bibinfo{booktitle}{\emph{{GRADES-NDA}}},
  \bibfield{editor}{\bibinfo{person}{Akhil Arora}, \bibinfo{person}{Semih
  Salihoglu}, {and} \bibinfo{person}{Nikolay Yakovets}} (Eds.).
  \bibinfo{publisher}{{ACM}}, \bibinfo{pages}{4:1--4:11}.
\newblock


\bibitem[Han et~al\mbox{.}(2019)]%
        {conf/sigmod/HanKGPH19}
\bibfield{author}{\bibinfo{person}{Myoungji Han}, \bibinfo{person}{Hyunjoon
  Kim}, \bibinfo{person}{Geonmo Gu}, \bibinfo{person}{Kunsoo Park}, {and}
  \bibinfo{person}{Wook{-}Shin Han}.} \bibinfo{year}{2019}\natexlab{}.
\newblock \showarticletitle{Efficient Subgraph Matching: Harmonizing Dynamic
  Programming, Adaptive Matching Order, and Failing Set Together}. In
  \bibinfo{booktitle}{\emph{{SIGMOD}}}. \bibinfo{publisher}{{ACM}},
  \bibinfo{pages}{1429--1446}.
\newblock


\bibitem[Han et~al\mbox{.}(2018)]%
        {conf/sigmod/Han0Y18}
\bibfield{author}{\bibinfo{person}{Shuo Han}, \bibinfo{person}{Lei Zou}, {and}
  \bibinfo{person}{Jeffrey~Xu Yu}.} \bibinfo{year}{2018}\natexlab{}.
\newblock \showarticletitle{Speeding Up Set Intersections in Graph Algorithms
  using {SIMD} Instructions}. In \bibinfo{booktitle}{\emph{{SIGMOD}}}.
  \bibinfo{publisher}{{ACM}}, \bibinfo{pages}{1587--1602}.
\newblock


\bibitem[Jin et~al\mbox{.}(2021)]%
        {conf/icde/JinY0YQP21}
\bibfield{author}{\bibinfo{person}{Xin Jin}, \bibinfo{person}{Zhengyi Yang},
  \bibinfo{person}{Xuemin Lin}, \bibinfo{person}{Shiyu Yang},
  \bibinfo{person}{Lu Qin}, {and} \bibinfo{person}{You Peng}.}
  \bibinfo{year}{2021}\natexlab{}.
\newblock \showarticletitle{{FAST:} FPGA-based Subgraph Matching on Massive
  Graphs}. In \bibinfo{booktitle}{\emph{37th {IEEE} International Conference on
  Data Engineering, {ICDE} 2021, Chania, Greece, April 19-22, 2021}}.
  \bibinfo{publisher}{{IEEE}}, \bibinfo{pages}{1452--1463}.
\newblock
\urldef\tempurl%
\url{https://doi.org/10.1109/ICDE51399.2021.00129}
\showDOI{\tempurl}


\bibitem[Lasch et~al\mbox{.}(2022)]%
        {conf/edbt/LaschMMDFS22}
\bibfield{author}{\bibinfo{person}{Robert Lasch}, \bibinfo{person}{Mehdi
  Moghaddamfar}, \bibinfo{person}{Norman May},
  \bibinfo{person}{S{\"{u}}leyman~Sirri Demirsoy}, \bibinfo{person}{Christian
  F{\"{a}}rber}, {and} \bibinfo{person}{Kai{-}Uwe Sattler}.}
  \bibinfo{year}{2022}\natexlab{}.
\newblock \showarticletitle{Bandwidth-optimal Relational Joins on FPGAs}. In
  \bibinfo{booktitle}{\emph{{EDBT}}}. \bibinfo{pages}{1:27--1:39}.
\newblock


\bibitem[Lee et~al\mbox{.}(2012)]%
        {journals/pvldb/LeeHKL12}
\bibfield{author}{\bibinfo{person}{Jinsoo Lee}, \bibinfo{person}{Wook{-}Shin
  Han}, \bibinfo{person}{Romans Kasperovics}, {and}
  \bibinfo{person}{Jeong{-}Hoon Lee}.} \bibinfo{year}{2012}\natexlab{}.
\newblock \showarticletitle{An In-depth Comparison of Subgraph Isomorphism
  Algorithms in Graph Databases}.
\newblock \bibinfo{journal}{\emph{Proc. {VLDB} Endow.}} \bibinfo{volume}{6},
  \bibinfo{number}{2} (\bibinfo{year}{2012}), \bibinfo{pages}{133--144}.
\newblock


\bibitem[Leskovec and Krevl(2014)]%
        {LeskovecK14}
\bibfield{author}{\bibinfo{person}{Jure Leskovec} {and} \bibinfo{person}{Andrej
  Krevl}.} \bibinfo{year}{2014}\natexlab{}.
\newblock \bibinfo{title}{{SNAP Datasets}: {Stanford} Large Network Dataset
  Collection}.
\newblock \bibinfo{howpublished}{\url{http://snap.stanford.edu/data}}.
\newblock


\bibitem[Mawhirter et~al\mbox{.}(2021)]%
        {journals/sigops/MawhirterRHLW21}
\bibfield{author}{\bibinfo{person}{Daniel Mawhirter}, \bibinfo{person}{Sam
  Reinehr}, \bibinfo{person}{Connor Holmes}, \bibinfo{person}{Tongping Liu},
  {and} \bibinfo{person}{Bo Wu}.} \bibinfo{year}{2021}\natexlab{}.
\newblock \showarticletitle{GraphZero: {A} High-Performance Subgraph Matching
  System}.
\newblock \bibinfo{journal}{\emph{{ACM} {SIGOPS} Oper. Syst. Rev.}}
  \bibinfo{volume}{55}, \bibinfo{number}{1} (\bibinfo{year}{2021}),
  \bibinfo{pages}{21--37}.
\newblock


\bibitem[Mhedhbi and Salihoglu(2019)]%
        {journals/pvldb/MhedhbiS19}
\bibfield{author}{\bibinfo{person}{Amine Mhedhbi} {and} \bibinfo{person}{Semih
  Salihoglu}.} \bibinfo{year}{2019}\natexlab{}.
\newblock \showarticletitle{Optimizing Subgraph Queries by Combining Binary and
  Worst-Case Optimal Joins}.
\newblock \bibinfo{journal}{\emph{{PVLDB}}} \bibinfo{volume}{12},
  \bibinfo{number}{11} (\bibinfo{year}{2019}), \bibinfo{pages}{1692--1704}.
\newblock


\bibitem[Ngo et~al\mbox{.}(2018)]%
        {journals/jacm/NgoPRR18}
\bibfield{author}{\bibinfo{person}{Hung~Q. Ngo}, \bibinfo{person}{Ely Porat},
  \bibinfo{person}{Christopher R{\'{e}}}, {and} \bibinfo{person}{Atri Rudra}.}
  \bibinfo{year}{2018}\natexlab{}.
\newblock \showarticletitle{Worst-case Optimal Join Algorithms}.
\newblock \bibinfo{journal}{\emph{J. {ACM}}} \bibinfo{volume}{65},
  \bibinfo{number}{3} (\bibinfo{year}{2018}), \bibinfo{pages}{16:1--16:40}.
\newblock


\bibitem[Ngo et~al\mbox{.}(2013)]%
        {journals/sigmod/NgoRR13}
\bibfield{author}{\bibinfo{person}{Hung~Q. Ngo}, \bibinfo{person}{Christopher
  R{\'{e}}}, {and} \bibinfo{person}{Atri Rudra}.}
  \bibinfo{year}{2013}\natexlab{}.
\newblock \showarticletitle{Skew strikes back: new developments in the theory
  of join algorithms}.
\newblock \bibinfo{journal}{\emph{{SIGMOD} Rec.}} \bibinfo{volume}{42},
  \bibinfo{number}{4} (\bibinfo{year}{2013}), \bibinfo{pages}{5--16}.
\newblock


\bibitem[Przulj et~al\mbox{.}(2006)]%
        {journals/bioinformatics/PrzuljCJ06}
\bibfield{author}{\bibinfo{person}{Natasa Przulj}, \bibinfo{person}{Derek~G.
  Corneil}, {and} \bibinfo{person}{Igor Jurisica}.}
  \bibinfo{year}{2006}\natexlab{}.
\newblock \showarticletitle{Efficient estimation of graphlet frequency
  distributions in protein-protein interaction networks}.
\newblock \bibinfo{journal}{\emph{Bioinform.}} \bibinfo{volume}{22},
  \bibinfo{number}{8} (\bibinfo{year}{2006}), \bibinfo{pages}{974--980}.
\newblock


\bibitem[Rossi and Ahmed(2015)]%
        {conf/aaai/RossiA15}
\bibfield{author}{\bibinfo{person}{Ryan~A. Rossi} {and}
  \bibinfo{person}{Nesreen~K. Ahmed}.} \bibinfo{year}{2015}\natexlab{}.
\newblock \showarticletitle{The Network Data Repository with Interactive Graph
  Analytics and Visualization}.
  \bibinfo{howpublished}{\url{http://networkrepository.com}}. In
  \bibinfo{booktitle}{\emph{AAAI}}.
\newblock


\bibitem[Sahu et~al\mbox{.}(2020)]%
        {journals/vldb/SahuMSLO20}
\bibfield{author}{\bibinfo{person}{Siddhartha Sahu}, \bibinfo{person}{Amine
  Mhedhbi}, \bibinfo{person}{Semih Salihoglu}, \bibinfo{person}{Jimmy Lin},
  {and} \bibinfo{person}{M.~Tamer {\"{O}}zsu}.}
  \bibinfo{year}{2020}\natexlab{}.
\newblock \showarticletitle{The ubiquity of large graphs and surprising
  challenges of graph processing: extended survey}.
\newblock \bibinfo{journal}{\emph{{VLDB} J.}} \bibinfo{volume}{29},
  \bibinfo{number}{2-3} (\bibinfo{year}{2020}), \bibinfo{pages}{595--618}.
\newblock


\bibitem[Snijders et~al\mbox{.}(2006)]%
        {journals/sociometh/SnijdersPRH06}
\bibfield{author}{\bibinfo{person}{Tom~AB Snijders},
  \bibinfo{person}{Philippa~E Pattison}, \bibinfo{person}{Garry~L Robins},
  {and} \bibinfo{person}{Mark~S Handcock}.} \bibinfo{year}{2006}\natexlab{}.
\newblock \showarticletitle{New specifications for exponential random graph
  models}.
\newblock \bibinfo{journal}{\emph{Sociological methodology}}
  \bibinfo{volume}{36}, \bibinfo{number}{1} (\bibinfo{year}{2006}),
  \bibinfo{pages}{99--153}.
\newblock


\bibitem[Sun and Luo(2020)]%
        {conf/sigmod/Sun020}
\bibfield{author}{\bibinfo{person}{Shixuan Sun} {and} \bibinfo{person}{Qiong
  Luo}.} \bibinfo{year}{2020}\natexlab{}.
\newblock \showarticletitle{In-Memory Subgraph Matching: An In-depth Study}. In
  \bibinfo{booktitle}{\emph{{SIGMOD}}}. \bibinfo{publisher}{{ACM}},
  \bibinfo{pages}{1083--1098}.
\newblock


\bibitem[Sun et~al\mbox{.}(2020)]%
        {journals/pvldb/SunSC0H20}
\bibfield{author}{\bibinfo{person}{Shixuan Sun}, \bibinfo{person}{Xibo Sun},
  \bibinfo{person}{Yulin Che}, \bibinfo{person}{Qiong Luo}, {and}
  \bibinfo{person}{Bingsheng He}.} \bibinfo{year}{2020}\natexlab{}.
\newblock \showarticletitle{RapidMatch: {A} Holistic Approach to Subgraph Query
  Processing}.
\newblock \bibinfo{journal}{\emph{Proc. {VLDB} Endow.}} \bibinfo{volume}{14},
  \bibinfo{number}{2} (\bibinfo{year}{2020}), \bibinfo{pages}{176--188}.
\newblock


\bibitem[Ullmann(1976)]%
        {journals/jacm/Ullmann76}
\bibfield{author}{\bibinfo{person}{Julian~R. Ullmann}.}
  \bibinfo{year}{1976}\natexlab{}.
\newblock \showarticletitle{An Algorithm for Subgraph Isomorphism}.
\newblock \bibinfo{journal}{\emph{J. {ACM}}} \bibinfo{volume}{23},
  \bibinfo{number}{1} (\bibinfo{year}{1976}), \bibinfo{pages}{31--42}.
\newblock


\bibitem[Veldhuizen(2012)]%
        {journals/corr/abs-1210-0481}
\bibfield{author}{\bibinfo{person}{Todd~L. Veldhuizen}.}
  \bibinfo{year}{2012}\natexlab{}.
\newblock \showarticletitle{Leapfrog Triejoin: a worst-case optimal join
  algorithm}.
\newblock \bibinfo{journal}{\emph{CoRR}}  \bibinfo{volume}{abs/1210.0481}
  (\bibinfo{year}{2012}).
\newblock


\end{thebibliography}

\end{document}